\documentclass[twoside,12pt]{article}
\usepackage{epsfig}

\newcommand{\be}{\begin{equation}}
\newcommand{\ee}{\end{equation}}
\newcommand{\bea}{\begin{eqnarray}}
\newcommand{\eea}{\end{eqnarray}}

\begin{document}

\title{ \vspace{1cm} UHECR besides $Cen_A$: hints of galactic sources
}
\author{Daniele Fargion$^1,^2$,  \\
$^1$INFN,Rome University1, Italy\\
$^2$ Physics Department,Rome University Sapienza\\
}
\maketitle
\begin{abstract} Ultra High Energy Cosmic Rays, UHECR, maybe protons, as most still believe and claim, or nuclei; in particular lightest nuclei
 as we advocated recently. The two model offer a dramatic different view of UHECR sky because different galactic Lorentz deflection and GZK cut-off. The first (Auger Collaboration) nucleon proposal (2007)\cite{Auger-Nov07} foresaw  to trace clearly the UHECR GZK Universe reaching far (up to $100$ Mpc) Super-Galactic-Plane, with little angular dispersion. On the contrary  Lightest Nuclei model (2008)\cite{Fargion2008}, inspired by observed composition and by nearest $Cen_A$  clustering (almost a quarter of the AUGER events) explains (by cut off)  a modest and narrow (few Mpc) Universe view, as well as the puzzling Virgo absence. Lightest nuclei offer a little blurred Astronomy.  Here we address to a part of the remaining scattered  events in the new up-dated Auger map (March 2009-ICRC09). We found within rarest clustering the surprising imprint of a few remarkable galactic sources. In particular we recognize a first trace of Vela, brightest gamma and radio galactic source, and smeared sources along Galactic Plane and Center. We expect in a near future much more clustering along $Cen_A$ and a few more confirm to those galactic sources. The clustering may imply additional tails of fragments (by nuclei photo-dissociation) at half energies ($2-4  \cdot 10^{19}$eV). The UHECR light-nuclei fragility and opacity may also reflect into a train of secondaries as gamma and neutrinos UHE events at tens-hundred PeVs. These UHE neutrinos might be detectable in a coming future within nearest AUGER and  Array Fluorescence Telescope views,(few $km$ distances) by fast fluorescence   flashing of horizontal up-going $\tau$ Air-showers.
\end{abstract}
\section{Introduction: Lorentz bending from $Cen_A$ and Vela}
 The study of UHECR map maybe correlated with different astronomy wave-band: optical, X, radio, gamma at different energies.
 The optical map fails to show any connection with UHECR. The Infrared map shows an uncorrelated and  missing of  Virgo bump. The nearest AGN catalog does not correlate much too.  The main correlated map is the $408$ MHz one. The first astronomical source that seem to correlate is the main multiplet along $Cen_A$.  This AGN source, the nearest extragalactic one,  sits in the same direction of a far Centaurus Cluster (part of the Super-Galactic Plane). The blurring by random galactic magnetic field might spread the nearest AGN event along the same Super-Galactic Plane, explaining the AUGER group miss-understanding \cite{Fargion2008}.  Composition derived by slant-depth $X_{max}$ points to light and maybe few heavy nuclei.
  The probable clustering toward $Cen_A$ favors lightest ones. As well as the absence of Virgo Cluster \cite{Gorbunov09}, \cite{Fargion2008}.
\begin{figure}[tb]
\begin{center}
\epsfig{file=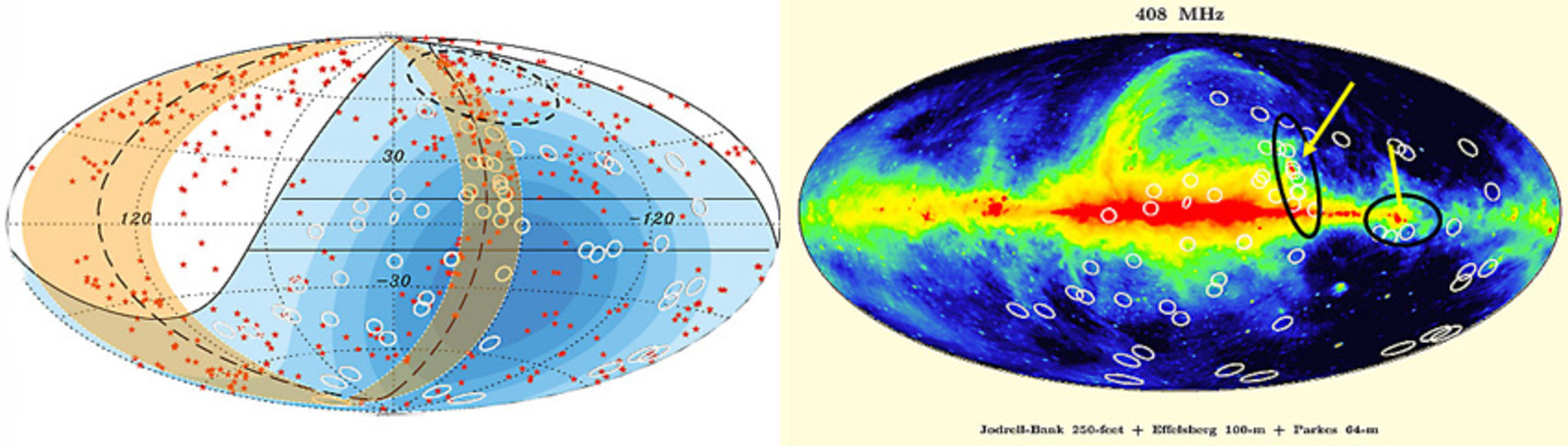,scale=0.47}
\begin{minipage}[t]{16.5 cm}
\caption{On the left the new AUGER map up to March $2009$ with $58$ events \cite{Scineghe09}.
 The shaded areas are correlated to the nearest GZK \cite{Greisen:1966jv} volumes, namely the Super Galactic Plane. Their apparent earliest correlation \cite{Auger-Nov07} partially faded away because of recent spread more isotropic UHECR events. The presence  of a  multiple along $Cen_A$ source  (almost a quarter of all the events)  was evident in old \cite{Fargion2008} and it is in new map in a persistent way. Note (dashed area)the puzzling Virgo absence. On the right side the same UHECR  map in radio $408$Mhz background. The brightest radio activity of nearest extra-galactic AGN,$Cen_A$,  as well as the brightest spot of near galactic gamma Pulsar Vela correlate with UHECR. The latter spot is well within the unique triplet (besides $Cen_A$). The probability that this unique clustering correlate by chance with Vela  is very small, well below one percent.
 }
\end{minipage}
\end{center}
\end{figure}
There are two main spectroscopy of UHECR along galactic plane:
 A late nearby (almost local) bending by a nearest coherent galactic arm field, and a random one along the whole plane.
The coherent Lorentz angle bending $\delta_{Coh} $ of a proton UHECR (above GZK \cite{Greisen:1966jv}) within a galactic magnetic field  in a final nearby coherent length  of $l_c = 1\cdot kpc$ is $ \delta_{Coh-p} \simeq{2.3^\circ}\cdot \frac{Z}{Z_{H}} \cdot (\frac{6\cdot10^{19}eV}{E_{CR}})(\frac{B}{3\cdot \mu G}){\frac{l_c}{kpc}}$.
The corresponding coherent  bending of an Helium UHECR at same energy, within a galactic magnetic field
  in a wider nearby coherent length  of $l_c = 2\cdot  kpc$ is
\begin{equation}
\delta_{Coh-He} \simeq
{9.2^\circ}\cdot \frac{Z}{Z_{He}} \cdot (\frac{6\cdot10^{19}eV}{E_{CR}})(\frac{B}{3\cdot \mu G}){\frac{l_c}{2 kpc}}
\end{equation}
This bending angle is compatible with observed triplet clustering along Vela, at much nearer distances, for a larger magnetic field along its direction (20 $\mu G$) or for rare iron composition $\delta_{Coh-Fe-Vela} \simeq
{17.4^\circ}\cdot \frac{Z}{Z_{Fe}} \cdot (\frac{6\cdot10^{19}eV}{E_{CR}})(\frac{B}{3\cdot \mu G}){\frac{l_c}{290 pc}}$. Such iron UHECR are mostly bounded inside the Galaxy, as well as in Virgo, explaining its absence. The heavier of lightest nuclei that may be bounded from Virgo, Be, is bent by $
\delta_{Coh-Be} \simeq
{18.4^\circ}\cdot \frac{Z}{Z_{Be}} \cdot (\frac{6\cdot10^{19}eV}{E_{CR}})(\frac{B}{3\cdot \mu G}){\frac{l_c}{2 kpc}}
$.  The incoherent random angle bending, $\delta_{rm} $, while crossing along the whole Galactic disk $ L\simeq{20 kpc}$  in different spiral arms  and within a characteristic  coherent length  $ l_c \simeq{2 kpc}$ for He nuclei is
$
\delta_{rm-He} \simeq
{16^\circ}\cdot \frac{Z}{Z_{He^2}} \cdot (\frac{6\cdot10^{19}eV}{E_{CR}})(\frac{B}{3\cdot \mu G})\sqrt{\frac{L}{20 kpc}}
\sqrt{\frac{l_c}{2 kpc}}
$
The heavier  (but still lightest nuclei) bounded from Virgo are Li and Be:
$\delta_{rm-Li} \simeq
{24^\circ}\cdot \frac{Z}{Z_{Li^3}} \cdot (\frac{6\cdot10^{19}eV}{E_{CR}})(\frac{B}{3\cdot \mu G})\sqrt{\frac{L}{20 kpc}}
\sqrt{\frac{l_c}{2 kpc}}
$, $
\delta_{rm-Be} \simeq
{32^\circ}\cdot \frac{Z}{Z_{Be^4}} \cdot (\frac{6\cdot10^{19}eV}{E_{CR}})(\frac{B}{3\cdot \mu G})\sqrt{\frac{L}{20 kpc}}
\sqrt{\frac{l_c}{2 kpc}}
$.  It should be noted that the present anisotropy above GZK \cite{Greisen:1966jv} energy $5.5 \cdot 10^{19} eV$ might leave a tail of signals: indeed the photo disruption of He into deuterium, Tritium, $He^3$ and protons (and unstable neutrons), might rise as clustered events at half or a fourth of the energy.  It is important to look for correlated tails of events, possibly in  strings at low $\simeq 1.5-3 \cdot 10^{19} eV$ along the $Cen_A$ train of events. In conclusion He like UHECR  maybe bent by a characteristic as large as  $\delta_{rm-He}  \simeq 16^\circ$. Well within the observed CenA UHECR clustering spread.
\begin{figure}[tb]
\begin{center}
\epsfig{file=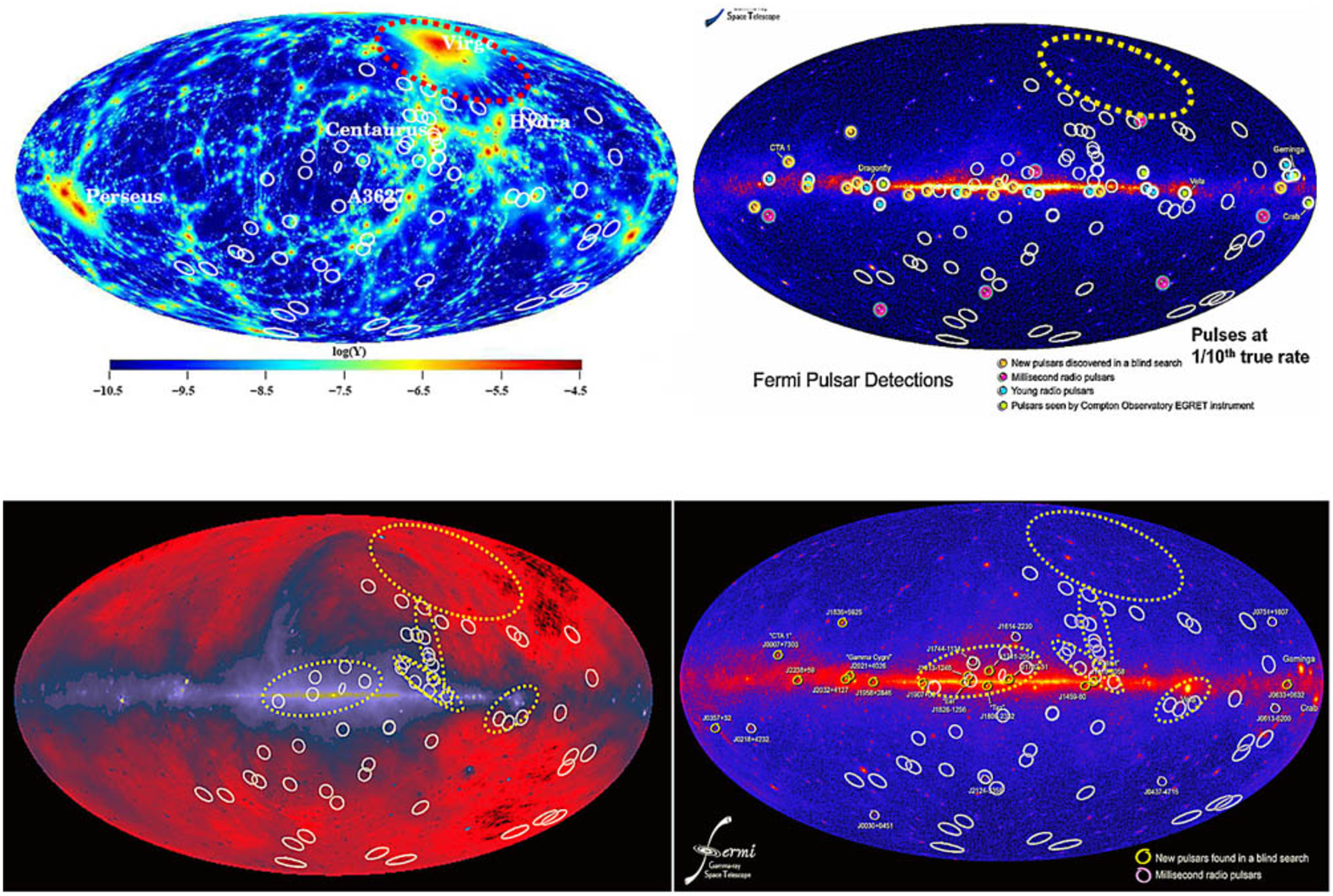,scale=0.42}
\begin{minipage}[t]{16.5 cm}
\caption{Top Left: Infrared Universe (Local Universe)  with the last AUGER UHECR $ 58 $ event rings. The absence of UHECR to the huge Virgo spot need an explanation. Lightest Nuclei fragility and consequent short (few Mpc) cut-off explains the Virgo absence or suppresion. Top right, the same UHECR over recent Fermi gamma map  and peculiar millisecond pulsars. The remarkable Vela shining both in gamma (hundreds MeV-GeV) and radio, as well as its unique triplet clustering well within the UHECR He or Fe bending angle, strongly suggest a Vela candidature as a first UHECR Galactic source. Down, left: other correlation of UHECR on Galactic plane,  shown over a radio map and shadow areas. Also Galactic center may shine in some form. Down right: similar clustering maps over different Fermi label sky. More discussion on other clustering will be considered elsewhere.
\label{fig1}}
\end{minipage}
\end{center}
\end{figure}

\emph{In conclusions} we are finally connecting UHECR sources by gamma ones and in future to UHE neutrino sky \cite{Fargion2008}, \cite{Auger08}.
The UHECR clustering on $Cen_A$ and a few along our galactic volumes open a less ambitious wide astronomy, but a nearest study of the most energetic cosmic ray accelerators: surprisingly both galactic and extra-galactic.
This article is devoted to the memory of great man and scientist \emph{Vitaly Ghinzburg}, ($October 4th, 1916 $-$November 8th, 2009$) who deceased just two weeks ago.

\begin{figure}[tb]
\begin{center}
\epsfig{file=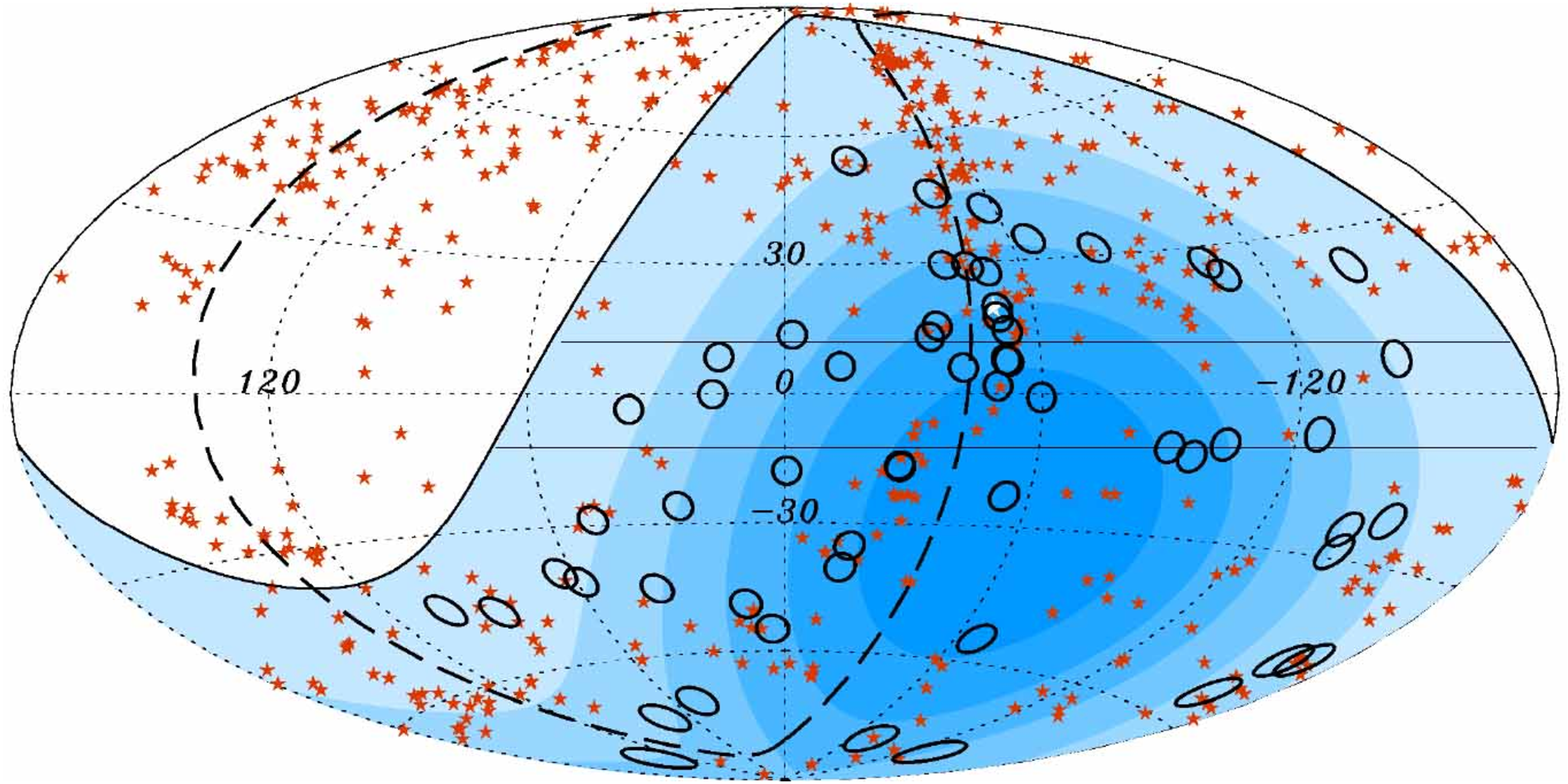,scale=0.220}
\begin{minipage}[t]{16.5 cm}
\caption{Original Slide 32 over 40 of D.Martello presentation on behalf of AUGER collaboration, on Scineghe VII Conference, Assisi October 2009. It should be noted that this presentation is, up to date, November 26th 2009, public and available on the web:
"http://agenda.infn.it/conferenceOtherViews.py?view=standard$\&$amp;confId=1369", Contribute.21.\cite{Scineghe09}
\label{fig1}}
\end{minipage}
\end{center}
\end{figure}

\bibliographystyle{elsarticle-num}
\bibliography{refs}

\end{document}